\documentclass[12pt,preprint,a4paper,superscriptaddress]{revtex4-1}
\usepackage{amsmath,amssymb}
\usepackage[utf8]{inputenc}
\usepackage{graphicx}
\usepackage{color}
\definecolor{mygray}{gray}{.4}
\usepackage{lineno}
\newcommand*\patchAmsMathEnvironmentForLineno[1]{%
\expandafter\let\csname old#1\expandafter\endcsname\csname #1\endcsname
\expandafter\let\csname oldend#1\expandafter\endcsname\csname end#1\endcsname
\renewenvironment{#1}%
{\linenomath\csname old#1\endcsname}%
{\csname oldend#1\endcsname\endlinenomath}}%
\newcommand*\patchBothAmsMathEnvironmentsForLineno[1]{%
\patchAmsMathEnvironmentForLineno{#1}%
\patchAmsMathEnvironmentForLineno{#1*}}%
\AtBeginDocument{%
\patchBothAmsMathEnvironmentsForLineno{equation}%
\patchBothAmsMathEnvironmentsForLineno{align}%
\patchBothAmsMathEnvironmentsForLineno{flalign}%
\patchBothAmsMathEnvironmentsForLineno{alignat}%
\patchBothAmsMathEnvironmentsForLineno{gather}%
\patchBothAmsMathEnvironmentsForLineno{multline}%
}

\begin{document}
\title{{\color{black} Continuum and Kinetic Simulations of the Neutral Gas Flow in an Industrial 
Physical Vapor Deposition Reactor}}
\author{Kirsten Bobzin}
\affiliation{RWTH Aachen University,
Faculty of Mechanical Engineering,
Surface Engineering Institute,
D-52056 Aachen, Germany}
\author{Ralf Peter Brinkmann}
\affiliation{Ruhr University Bochum,
Faculty of Electrical Engineering and Information Technology,
Institute of Theoretical Electrical Engineering,
D-44780 Bochum, Germany}
\author{Thomas Mussenbrock}
\affiliation{Ruhr University Bochum,
Faculty of Electrical Engineering and Information Technology,
Institute of Theoretical Electrical Engineering,
D-44780 Bochum, Germany}
\author{Nazlim Bagcivan}
\affiliation{RWTH Aachen University,
Faculty of Mechanical Engineering,
Surface Engineering Institute,
D-52056 Aachen, Germany}
\author{Ricardo Henrique Brugnara}
\affiliation{RWTH Aachen University,
Faculty of Mechanical Engineering,
Surface Engineering Institute,
D-52056 Aachen, Germany}
\author{Marcel Sch\"afer}
\affiliation{RWTH Aachen University,
Faculty of Mechanical Engineering,
Surface Engineering Institute,
D-52056 Aachen, Germany}
\author{Jan Trieschmann}
\email{jan.trieschmann@rub.de}
\thanks{Corresponding author}
\affiliation{RWTH Aachen University,
Faculty of Mechanical Engineering,
Surface Engineering Institute,
D-52056 Aachen, Germany}
\affiliation{Ruhr University Bochum,
Faculty of Electrical Engineering and Information Technology,
Institute of Theoretical Electrical Engineering,
D-44780 Bochum, Germany}
\date{\today}

\begin{abstract}
Magnetron sputtering {\color{black} used for physical vapor 
deposition processes often requires} gas pressures well 
below 1 Pa. Under these conditions the gas flow in 
the reactor is usually determined by a Knudsen number
of about one, i.e., a transition regime between the 
hydrodynamic and the rarefied gas regime. In the first, 
the gas flow is well described by the Navier-Stokes 
equations, while in the second a kinetic approach
via the Boltzmann equation is necessary. 
{\color{black} In this paper the neutral gas flow of 
argon and molecular nitrogen gas inside an industrial 
scale plasma reactor was simulated using both a fluid model
and a fully kinetic Direct Simulation Monte Carlo model.
By comparing both model results the validity of the
fluid model was checked. Although in both models a 
Maxwell-Boltzmann energy distribution of the neutral
particles is the natural outcome, the results of 
the gas flow differ significantly. The fluid model 
description breaks down, due to the inappropriate
assumption of a fluid continuum. This is due 
to exclusion of non-local effects in the multi 
dimensional velocity space, as well as invalid gas/wall 
interactions. Only the kinetic model is able to 
provide an accurate physical description of the gas 
flow in the transition regime. Our analysis is 
completed with a brief investigation of different
definitions of the local Knudsen number. We conclude 
that the most decisive parameter -- the spatial 
length scale $L$ -- has to be very careful chosen in 
order to obtain a reasonable estimate of the gas 
flow regime.}
\end{abstract}

\maketitle

\newpage

\section{Introduction}\label{sec:introduction}

Physical Vapor Deposition (PVD) processes, such as the 
well established {\color{black} cathodic arc evaporation (CAE) and} DC Magnetron Sputtering (DC-MS),
{\color{black} as well as the promising} High Power {\color{black}Pulsed} Magnetron Sputtering 
(HPPMS, often referred to as HiPIMS) {\color{black} 
technology can be} used for the production of hard 
protective coatings in corrosion and wear resistance 
applications {\color{black}\cite{bobzin2008, theiss2010,
sanders2000}}. In this context a uniform layer of 
coating material is an essential requirement. To optimize
industrial PVD processes in terms of the quality of the 
obtained coatings, it is important to understand not only 
the discharge characteristics, but moreover to obtain a 
detailed picture of {\color{black} the neutral gas flow} 
inside the reactor chamber.

The governing parameter commonly used for the analysis
and the characterization of the gas flow regime is the 
Knudsen number $\textit{Kn}$ \cite{bird1983, bird1994}.
$\textit{Kn}$ allows to approximately estimate the flow 
regime in a given setup by specifying the degree of gas
rarefaction \cite{bird1983}. It is commonly defined as 
the ratio of the mean free path $\lambda$ to a {\color{black}
representative (but local)} spatial scale $L$,
\begin{gather}
\textit{Kn}=\frac{\lambda}{L}.\label{eq:knudsenNumber}
\end{gather}
{\color{black} In a gas in an equilibrium state with number 
density $n$, the mean free path can be estimated by
$\lambda=\left(n\sigma\sqrt{2}\right)^{-1}$, where the 
hard sphere collision cross section $\sigma=\pi d^2$ may 
be used. A representative length scale $L$ can be chosen 
based on geometric considerations. Moreover, following Boyd 
et al. \cite{boyd1995} it can be defined through the 
normalized gradient of a local flow property $Q$ by 
$L = |Q|/|\nabla Q|$. Bird suggests a choice of $Q$ based 
on the mass density $Q=\rho=mn$ \cite{bird1994}. This is 
unfeasible for incompressible flows, which can 
be assumed in the case investigated (with a low Mach number 
of at maximum $M = |\vec{v}|/c \approx 0.15$). Thus we choose 
the momentum as the characteristic flow property 
$Q = \rho \vec{v}$. This choice of \textit{Kn} based on either 
the geometric or the gradient approach is addressed in a 
later context.

Regarding the analysis of the flow regime based on 
$\textit{Kn}$, the following distinctions can be made:} 
for {\color{black} $\textit{Kn} \leq 0.1$}, continuum models 
based on the Navier-Stokes equations -- usually implemented 
in computational fluid dynamic (CFD) simulations -- allow 
for a precise description of the gas flow. Such CFD models 
have found widespread applications in manifold areas of 
aerospace and automotive engineering \cite{anderson1995,
ferziger2002}. In contrast, in {\color{black} situations 
where $\textit{Kn}>0.1$} the Navier-Stokes equations
prove inadequate for the description of rarefied gas flows, 
e.g., {\color{black} in micro/nano scale gas flows
\cite{darbandi2011, white2013}, or in gas flows commonly 
used in low pressure PVD applications. This is due to 
non-local effects in the multi dimensional velocity space, 
as well as inappropriate treatment of gas/wall interactions.
Moreover, in plasma processing a continuum representation
(implying a Maxwell-Boltzmann energy distribution) is likely 
not valid, due to the interaction of neutral gas particles 
with non-equilibrium species (heavy particle, or 
electrons), leading to an overall non-equilibrium 
situation}. Under {\color{black} rarefied conditions, in 
general,} only kinetic models based on
the Boltzmann equation provide an accurate description. 
Such scenarios are often solved by means of the Direct 
Simulation Monte Carlo (DSMC) method proposed by 
Bird \cite{bird1994}. {\color{black} At very low pressures 
where a stochastic description of the particle interaction 
with the background gas is justified}, the Test Particle 
Monte Carlo (TPMC) method \cite{bird1978} is most commonly 
used. {\color{black} When using the DSMC method the} Boltzmann equation is directly solved by means 
of following the trajectory of a sufficiently large number 
of pseudo-particles subject to collisions among themselves, 
as well as with the surrounding walls. It is interesting 
to note that, although hydrodynamic and Monte Carlo 
methods have been studied extensively \cite{sazhin1997, 
sun2004, zheng2006, wu2006, kolobov2007}, generalized 
limitations for the validity of the different models in 
the transition regime $0.01<\textit{Kn}<2$ are not given.

The aim of this work is to discuss the validity of a 
conventional continuum model in the transition regime. 
For the analysis we apply the commercially available 
continuum fluid solver \textit{FLUENT} \cite{fluent2012}. 
In order to allow for a detailed comparison, we have 
modified -- with respect to boundary conditions -- the 
DSMC solver \textit{dsmcFoam} \cite{scanlon2010, 
macpherson2009, openfoam2012}. In this work we analyze 
the neutral gas flow inside a reactor chamber used for 
DC-MS and HPPMS processes. We present a brief description 
of the investigated reactor system and motivate our 
analysis of the two different numerical algorithms. 
We give a short review of both numerical models. A 
detailed discussion of simulation results of the identical 
vacuum setup obtained via the two models is provided. 
Finally, the results are summarized and a conclusion is 
drawn. We provide suggestions on the validity of either 
continuum solution based CFD simulations and kinetic models.

\section{Setup}\label{sec:setup}

\begin{figure}[t]
\centering
\includegraphics[width=\textwidth]{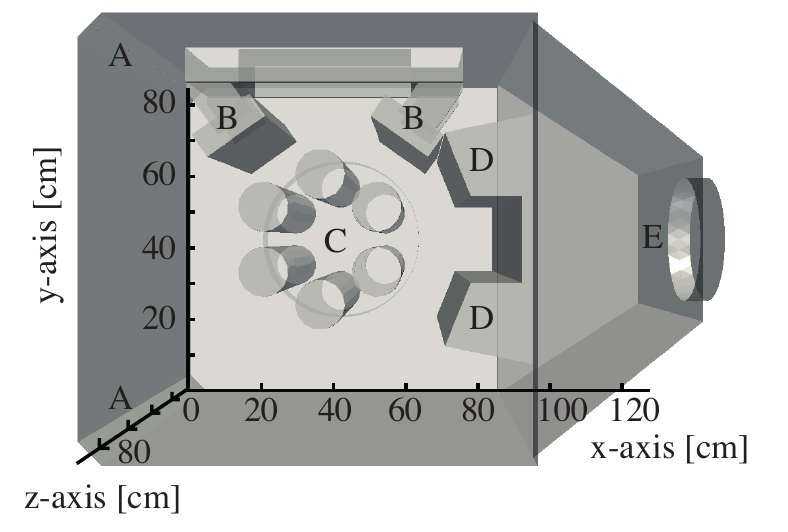}
\caption[3D top view of the PVD coating unit CC800/9 Custom. 
Gas inlets (A) in both corners at the far left side, pump 
flange (E) at far right side. Two cathodes (B) on one side 
of the substrate holders (C). The heater (D) separates the 
main from the pump chamber.]{}
\label{fig:3d_geometry}
\end{figure}

The CemeCon CC800/9 Custom coating unit, investigated in 
this work, is typically used for DC-MS and HPPMS processes. 
As illustrated in figure \ref{fig:3d_geometry}, 
{\color{black} the main processing chamber has a floor space 
of $85\times85$ cm$^2$.} Additionally, the main chamber 
is extended towards the pump by a narrowing pump chamber. 
While the gas inlets (A) are along two of the corners 
of the main chamber at the left side, the pump (E) is 
mounted at the large flange at the right side. Inside 
the main chamber there are six sample holders (C). These 
sample holders can be static or rotating using a 
planetary gearing. Behind the sample holders two 
magnetron cathodes (B) are mounted. The main chamber 
and the pump chamber are separated by a heater (D). The 
coating unit is typically operated at pressures around 
0.5 Pa. {\color{black} For this study we assume that argon} is used as process gas at a flow rate 
of $F_{Ar}=200$ sccm, while (molecular) nitrogen is used 
as reactive gas at a flow rate of $F_{N_2}=40$ sccm. All 
walls are assumed to have a constant temperature of 
$T=300$ K.

{\color{black} From geometric considerations, the typical 
geometric dimension $L$ ranges from about one centimeter 
(e.g., at small features,} the substrates and cathodes)
up to a few tens of centimeters (at open space in the 
vacuum chamber). {\color{black} Additionally, for the pressure
of 500 mPa, a mean free path of $\lambda \approx 1.39$ cm 
can be approximated for an equilibrium gas assuming a 
molecular diameter of $d=3.664$ \r{A} for argon
\cite{chapman1952}. In consequence, one finds the Knudsen 
number $\textit{Kn}=\lambda / L$ in the limits
$0.05<\textit{Kn}<1.5$.} While the lower limit suggests 
that continuum models can be readily used for a numerical
analysis, the upper limit enforces that only a kinetic 
treatment of particles is valid for a description of 
the {\color{black} gas flow}. In the transition regime, 
however, no definite statement can be made. Therefore, 
to gain insight in the gas flow -- and the models -- 
{\color{black} an investigation of results from a} continuum 
model based on the Navier-Stokes equations, as well as
{\color{black} a kinetic DSMC simulation model}, is desired.

\section{Numerical Models}\label{sec:numerical_models}

The modeling of magnetron sputtering processes, involving 
the interaction of plasma with {\color{black} 
neutral background gas (in this work argon and 
molecular nitrogen),} as well as the interaction of 
heavy particles with target materials (sputtering) 
and walls/substrates (deposition) is of very complex 
nature. Several authors have investigated the 
theoretical background of sputtering processes in 
terms of the plasma/wall interaction {\color{black}
\cite{thompson1968, sigmund1969, mussenbrock2012}}, the description of 
the deposition of sputtered material on substrates and 
walls {\color{black} (including the chemical interaction with 
gas phase species) \cite{berg1987, berg2005, depla2007}}, 
as well as the analysis of the plasma, particularly for 
HPPMS processes \cite{kouznetsov1999, lundin2009, 
sarakinos2010, gudmundsson2012}. 
The {\color{black} numerical} investigations are {\color{black} often} based on particle based 
models {\color{black} \cite{kersch1994, serikov1996, 
malaurie1996, lugscheider2005, mahieu2006, kadlec2007,
aeken2008, lundin2013}}. In this work, we concentrate 
on the {\color{black} neutral gas flow}; the interaction 
of the neutral gas with charged particles {\color{black} 
from} the plasma, {\color{black} energetic heavy 
particles sputtered off the targets (e.g., gas rarefaction effects)
\cite{huo2012}, as well as the resulting} 
interaction with the walls {\color{black} (i.e., the deposition process)} are intentionally 
left for a later analysis. For our analysis we employ 
the CFD software \textit{FLUENT} and a modified version 
of the DSMC implementation \textit{dsmcFoam} provided 
with the freely available \textit{OpenFOAM} simulation 
package \cite{openfoam2012}.

For the CFD simulations, the \textit{FLUENT} software 
release 14 \cite{fluent2012} is used. We simulate 
{\color{black} the gas flow using a pressure based fluid 
model using the PISO method 
\cite{ferziger2002}. Additional we use a RNG k-epsilon 
turbulence model \cite{yakhot1992} for robust 
convergence. For further improvement of} the 
numerical solution {\color{black} scheme}, two neighbor 
and two skewness correction iterations are applied, 
respectively. The ideal gas law is used to obtain the 
gas density from the pressure and the gas temperature. 
The walls are described by slip boundary conditions. 
As for the parameters, we set a mass flux and the 
wall temperature corresponding to the flow rate and 
temperature specified in the setup description. 
We further set the boundary condition at the pump to 
$p=430$ mPa.

The DSMC method is based on the idea that a sufficiently 
large number of pseudo-particles (also referred to as 
simulators) is kinetically simulated, interacting among 
each other by means of a given set of collision 
processes \cite{bird1994}. In this {\color{black} ensemble} 
(here in the converged state approximately 8 million 
simulators), each pseudo-particle represents a large 
number of physical particles, in our case $10^{13}$. 
By the original authors of \textit{dsmcFoam}, the 
solver was benchmarked against a number of examples from 
the literature {\color{black} \cite{scanlon2010}} and was 
used for various studies of rarefied gas flow {\color{black}\cite{white2013, darbandi2011}}.
In version 2.1.1 
used for this analysis, the implementation provides 
a simulation tool with the capabilities for arbitrary 
2D/3D geometries, an arbitrary number of gas species, 
variable hard sphere (VHS) collisions, Larsen-Borgnakke 
internal energy redistribution, and {\color{black} it} 
allows for unlimited parallel processing 
\cite{scanlon2010, bird1994}. The original code has 
been modified to allow for appropriate boundary conditions 
for {\color{black} both gas species in} the present 
problem: at the gas inlets {\color{black} mass flow 
rates are} specified and at the outlet{\color{black}/pump 
an absorption probability allows} to indirectly 
assign the pressure inside the vacuum chamber. 
{\color{black} If a particle hits the pump surface, 
with a probability $\alpha_\text{abs}$ it is 
removed from the simulation domain; otherwise it is 
thermally re-emitted into the volume. This procedure 
is performed for both species, argon and molecular 
nitrogen, respectively. In consequence}, the modified 
version of \textit{dsmcFoam} allows to simulate the 
neutral gas flow inside arbitrary bounded geometries. 
Regarding the specified parameters, we again assume 
a mass flux and wall temperature as presented above. 
Further we set the absorption rate at the pump to 
{\color{black} a pressure-fitted} value of 
$\alpha_\text{abs} = 17.25\%$. {\color{black} For the
VHS collision cross sections we use the values 
suggested by \cite{bird1994}. The Larsen-Borgnakke 
scheme for redistribution of internal energies is 
applied only to species with internal degrees of 
freedom available, which in our case -- due to its 
diatomic nature -- does affect only the molecular nitrogen.}
In addition, we assume complete thermalization 
of the particles once they impinge on the walls. To 
accurately describe the temporal and spatial dynamics, 
we apply a time-step size of $\Delta t = 25$ $\mu$s 
(a fraction of the mean collision time) and a typical 
cell size of about $\Delta x \approx 1 - 2$ cm 
{\color{black} (at maximum approximately the mean free path)}.

Due to the fundamental ``ab-initio'' treatment a 
DSMC model describes the gas flow in the transition regime
more accurately compared to CFD models. This holds 
particularly for particle/wall interactions. This 
hypothesis will be verified in the next section, 
where we present 3D simulation results for the CC800/9 
Custom coating unit. For both numerical approaches the 
same geometry has been used.

\section{Results and Discussion}\label{sec:results}

\begin{figure}[t]
\centering
\includegraphics[width=0.9\textwidth]{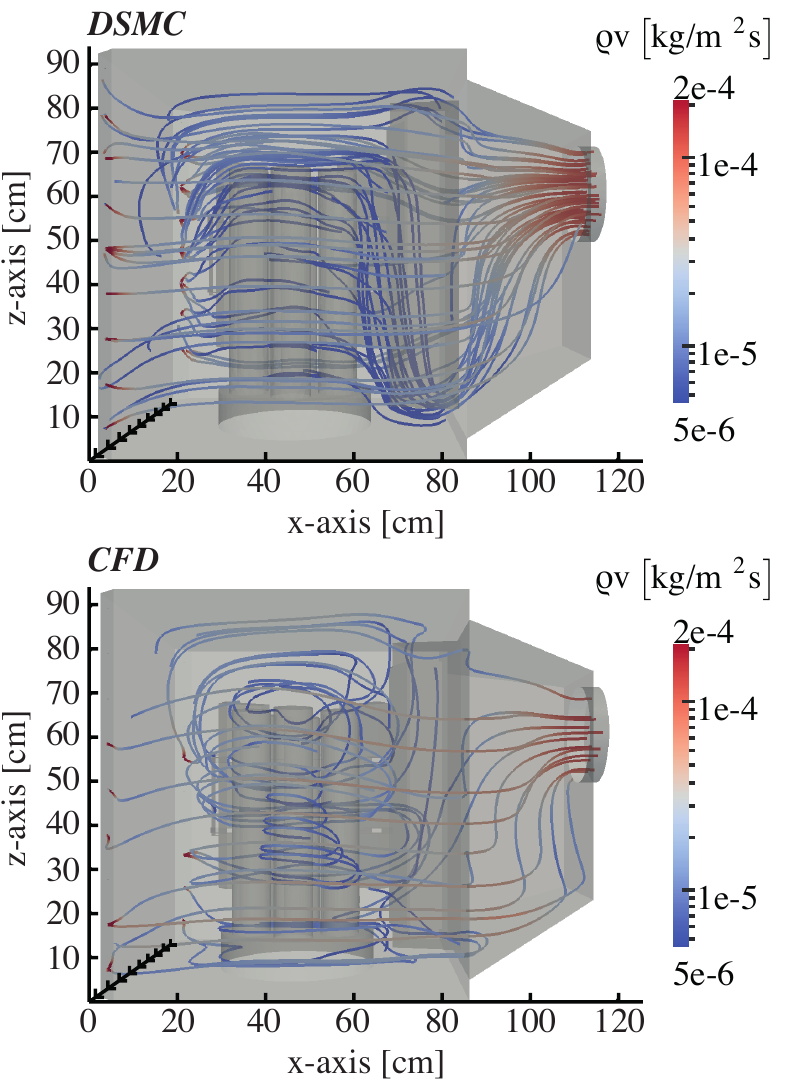}
\caption[Streamlines of the mass flux (or 
momentum) {\color{black}$\rho \vec{v}$} 
inside the coating unit (top: DSMC, bottom: CFD). Here, 
$\rho$ is the mass density and {\color{black}$\vec{v}$} 
is the average velocity.]{}
\label{fig:3d_side}
\end{figure}

We start our analysis of the two different model 
approaches by comparing the spatial distribution 
of the mass flux (or momentum) calculated by the 
two models. In figure \ref{fig:3d_side} the mass 
flux for both cases is indicated by the color-scaled 
streamlines. {\color{black} The quantitative} agreement
between both {\color{black} model results} is well
justified, as the mass flux through the chamber 
{\color{black} -- in both cases -- is} imposed by the 
boundary conditions. The absolute mass flow integrated 
over the areas of the inlet or outlet, respectively, 
is enforced to be exactly equal in both simulations. 
{\color{black}However, there is} a substantial difference in the 
spatial distribution of the mass flow{\color{black}. 
In the CFD results a large ``eddy'' appears, circulating 
around the substrate holders. It can be argued 
that this circulation may be a physical phenomenon. 
As we investigate a steady state result it is not an 
immediate consequence of the turbulence model which has been
used. Furthermore, in comparison with the more fundamental 
DSMC simulations (which naturally include turbulence) 
it is clearly not observed. In principle one can well 
imagine that there are scenarios that involve large 
circulating flows. However in our case the kinetic 
model does exclude this possibility. In summary we conclude that a break
down of the CFD model assumptions is observed.} 

\begin{figure}[t]
\centering
\includegraphics[width=0.7\textwidth]{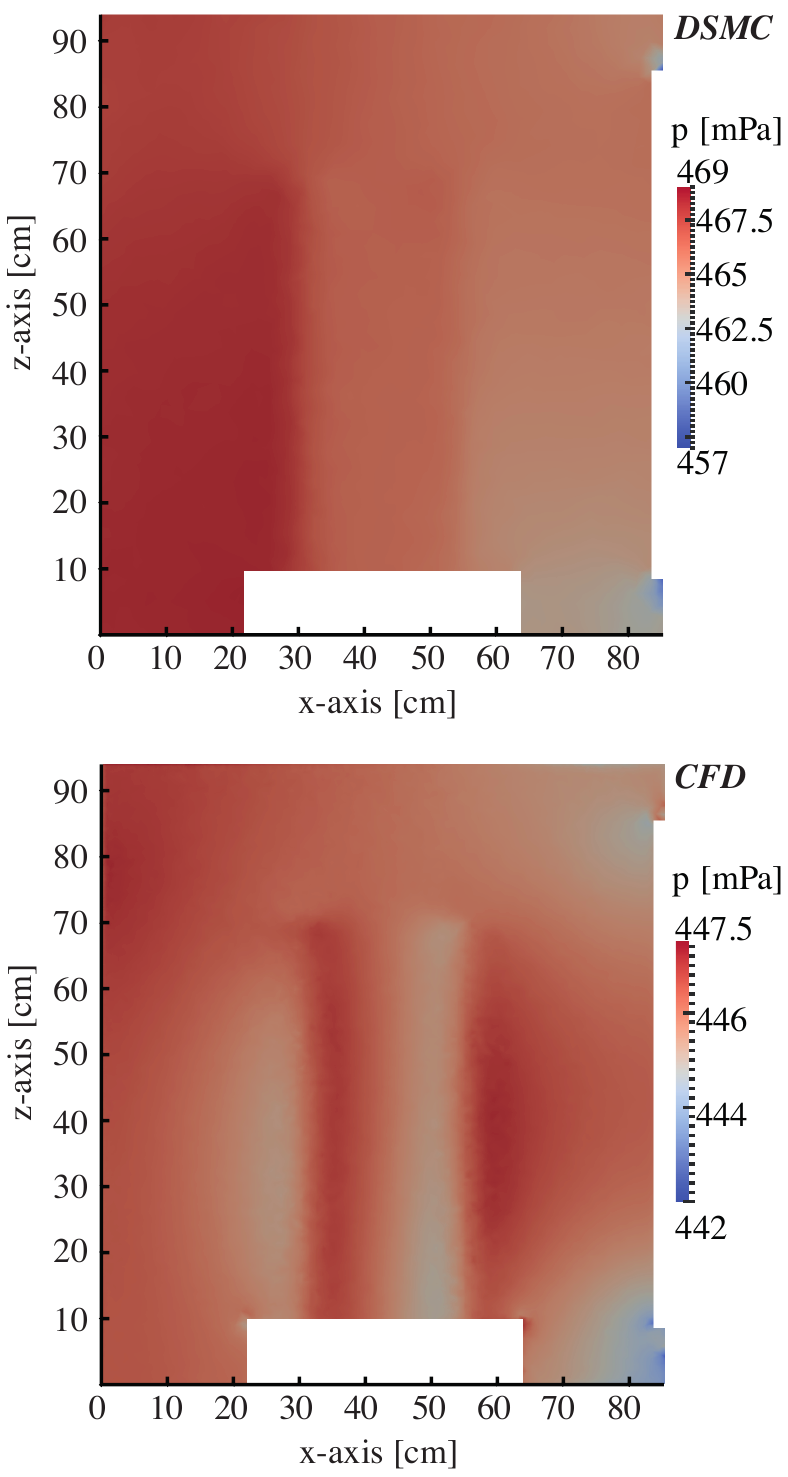}
\caption[Pressure profile of a cut along the $x$-axis through 
the center of the main vacuum chamber (top: DSMC, bottom: 
CFD).]{}
\label{fig:2d_center}
\end{figure}

Due to the externally imposed behavior of the total mass 
flow through the chamber, the spatial distribution of 
the pressure is a more reliable indicator for 
{\color{black} our} analysis. By considering a sectional 
plane along the direction of the $x$-axis, but in the 
center of the main vacuum chamber, a more detailed 
picture can be obtained. This is given in figure
\ref{fig:2d_center}. In the pressure profile 
calculated using the DSMC method a smooth decrease can 
be observed above the substrate holders {\color{black} 
($z>70$ cm)}. This region is characterized by open 
space and thus the flow is governed by a 
{\color{black} locally} small Knudsen number 
{\color{black} of $0.05<\textit{Kn}<0.15$}. This suggests 
that the CFD method can predict a similar -- and therefore
physical -- gas flow. {\color{black} {\color{black} Indeed, 
in} the CFD results (figure \ref{fig:2d_center}, bottom) 
the pressure drop in this region is also characterized by a smooth 
transition from the gas inlets (left) towards the 
pump (right).}

\begin{figure}[t]
\centering
\includegraphics[width=\textwidth]{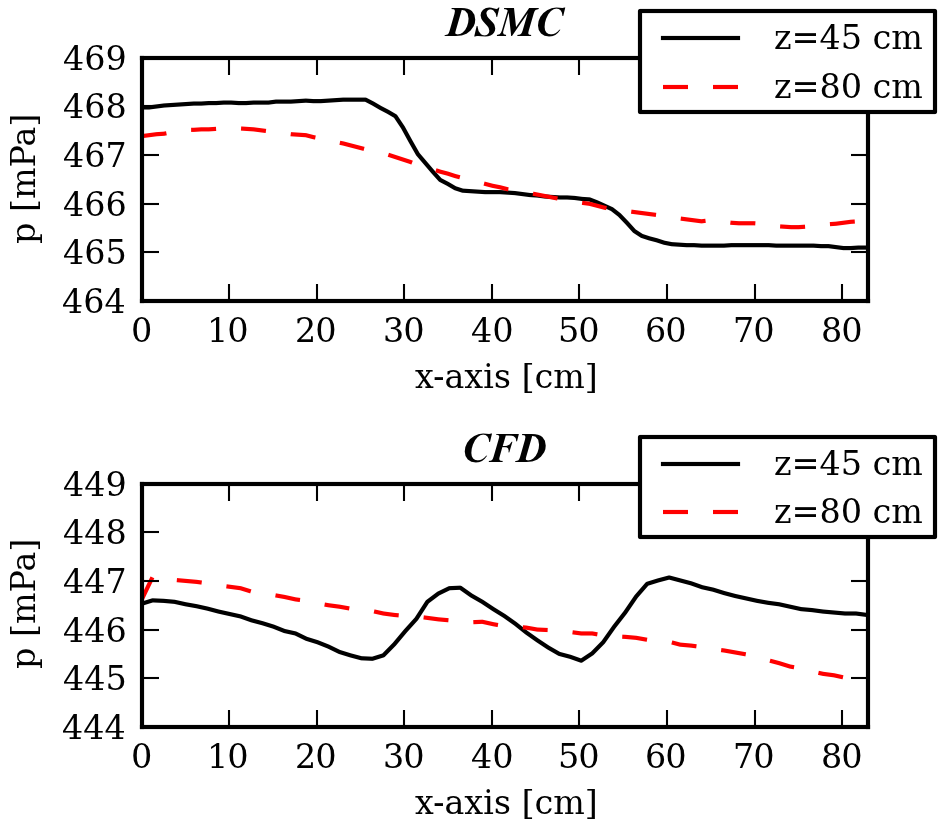}
\caption[Pressure inside the main chamber plotted along lines 
at different altitudes in direction of the $x$-axis (solid black: 
$z=45$ cm, dashed red: $z=80$ cm).]{}
\label{fig:1d_center}
\end{figure}

In contrast, in the center region at the substrate 
holders {\color{black} ($z<70$ cm)}, the DSMC {\color{black} 
results show} a more distinct, but still smooth pressure 
drop. In this more ``crowded'' region the Knudsen number 
is significantly larger, {\color{black} $0.3<\textit{Kn}<1$. 
A dynamic pressure occurs in front of the substrate holders 
and the flow bends around the obstacles.} Here the gas 
flow is dominated by {\color{black} particle/wall} 
interactions. The results obtained by the CFD model 
show an obviously different pressure profile. 
Strongly fluctuating -- and most likely unphysical --  
gradients in the pressure profile can be found between 
the far left wall at the gas inlets and the substrate 
holders. {\color{black} It can be reasoned that these 
fluctuations are a direct consequence of the large circulating flow 
observed for the mass flux. On the basis of the 
arguments presented earlier, however, a} physical 
explanation of these structures would be doubtful. 
This result suggests that, as expected, 
in regions of higher Knudsen numbers the CFD model 
reaches the limits of its validity. By investigating 
figure \ref{fig:1d_center}, where the pressure profiles 
for both approaches are exemplary extracted along two 
lines (in the central plane above and between the 
substrate holders), this becomes even more evident. 
While the DSMC results (figure \ref{fig:1d_center}, top) 
again suggest a smooth transition to lower pressures 
from the gas inlets towards the pump, the CFD 
results {\color{black} contradictory} claim a smooth, 
but strongly fluctuating pressure profile. This emphasizes 
the unphysical nature of {\color{black} the} CFD results in regions of larger 
Knudsen numbers.

\begin{figure}[t]
\centering
\includegraphics[width=0.7\textwidth]{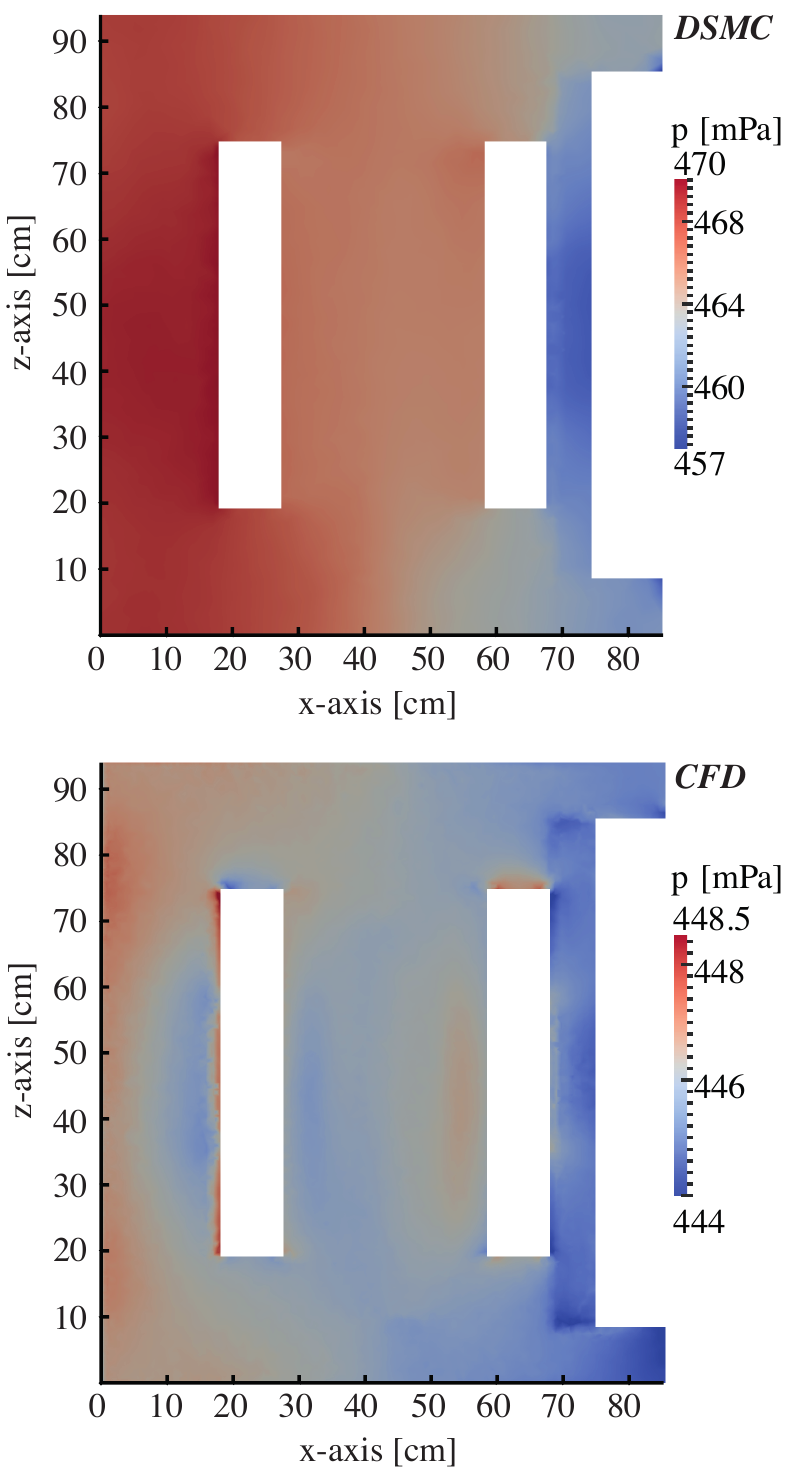}
\caption[Pressure profile of a cut in direction of 
the $x$-axis through the main chamber in the cathode 
plane (top: DSMC, bottom: CFD).]{}
\label{fig:2d_target}
\end{figure}

The results suggest that in regions where gas/wall 
interactions dominate, only kinetic models can 
accurately predict a physically correct gas flow. Even 
more pronounced, this is observed directly at wall 
surfaces, where the gas flow is {\color{black} strongly} distorted. 
In figure \ref{fig:2d_target}, a plane parallel to 
the sectional plane of figure \ref{fig:2d_center} is 
shown -- but now at the cathodes. Clearly, the effect 
of {\color{black} an incorrect treatment of the boundaries 
can be observed in the CFD simulations}. At small 
features, e.g., at the edges of the substrate holders, 
cathodes, or at the transition of the main to the pump 
chamber, numerical artifacts can be found. Here the 
CFD model predicts steep gradients (at the cathode 
surfaces) and layer-like structures (at the top and
bottom, far right in the main vacuum chamber). In 
contrast, the DSMC results show physically smooth 
pressure gradients. {\color{black} This proves} that
the CFD model is well justified in regions of small
and moderate Knudsen numbers, but fails in regions 
where the gas flow is dominated by gas/wall interactions.

An interesting aspect can be observed when investigating
the energy distribution of a given gas species in the 
kinetic simulations and the Maxwell-Boltzmann (MB) 
energy distribution, which is implicit in CFD models
\cite{anderson1995, reif1987}. In the latter, the 
temperature is a direct quantitative measure of the 
energy distribution. In contrast, in kinetic models 
like the particle based DSMC method the energy 
distribution of a gas species can in principle have 
any shape. {\color{black} This is the main advantage 
of kinetic models.} The shape is governed by all kinds 
of energy sources and/or sinks, e.g., due to the 
interaction of neutral particles with energetic 
particles or due to energy accommodation at walls 
of different temperatures.

For the modeling and simulation of magnetron
sputtering processes 
energy sources and sinks can be realized in terms of 
additional models {\color{black}(e.g., models of the plasma), 
or by inclusion of a flux of energetic non-equilibrium 
particles (e.g., sputtered particles)}. In this work, 
however, we concentrate on the neutral gas dynamics. 
(The coupling to a plasma model {\color{black} or the 
interaction with sputtered particles} will be part
of a subsequent study.) 

\begin{figure}[t]
\centering
\includegraphics[width=\textwidth]{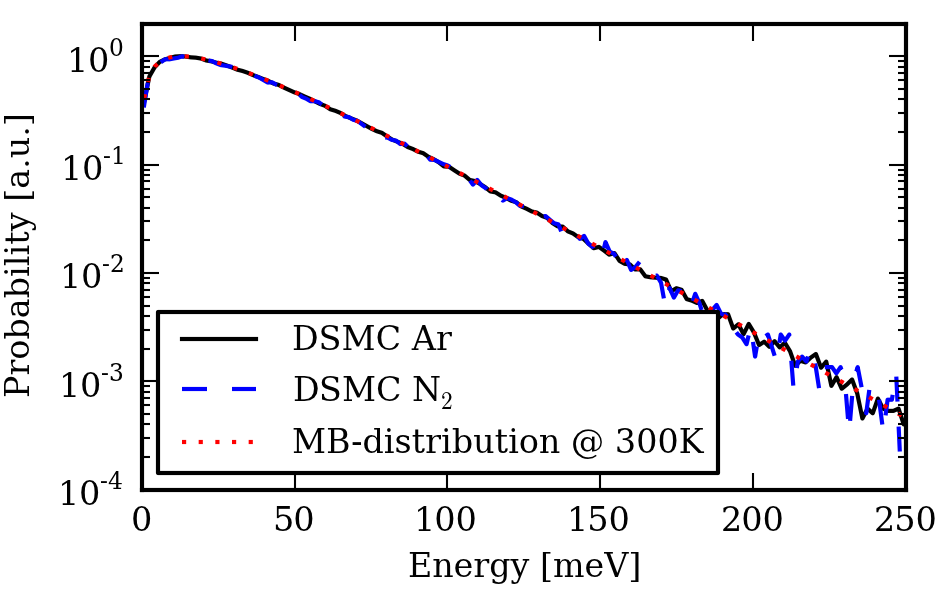}
\caption[Energy distribution function of argon and nitrogen gas 
species extracted from the DSMC simulations, as well as a plot 
of the Maxwell-Boltzmann distribution function for $T=300$ K. 
(The maximum is expected at 
$E_\text{max} = k_B T / 2 \approx 12.9$ meV).]{}
\label{fig:0d_particleDistribution}
\end{figure}

\begin{figure}[t]
\centering
\includegraphics[width=\textwidth]{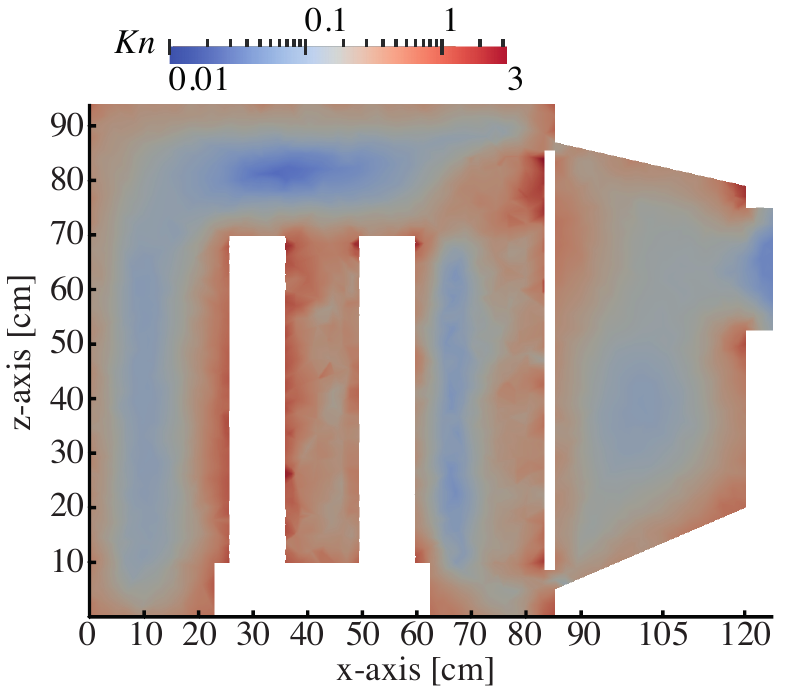}
\caption[{\color{black} Local Knudsen number based on 
equation (\ref{eq:knudsenNumber}) plotted along a cut 
through the main chamber in the plane of the substrate 
holders in direction of the $x$-axis. $\lambda=1.39$ cm 
is assumed. $L$ is given by the local momentum gradient 
as stated in section \ref{sec:introduction}.}]{}
\label{fig:2d_knudsenNumber}
\end{figure}

As suggested by the Boltzmann H-theorem \cite{reif1987}, 
for a neutral gas flow without additional energy 
sources/sinks a MB energy distribution is the natural 
outcome. In figure \ref{fig:0d_particleDistribution} the 
energy distributions of Ar and N$_2$ particles 
sampled from the DSMC model in the center region at 
the substrate holders are given. The results are 
compared to an analytical evaluation of the MB distribution 
for a defined temperature of 300 K. {\color{black} The} 
outcome of the fully kinetic simulations is in fact 
almost perfectly described by a MB energy distribution. 
Due to disagreement of the results for macroscopic
observables -- the spatial particle distribution and
the gas flow -- in the CFD and the DSMC model, it can
not be stated that the agreement of the energy 
distributions implies the validity of continuum 
methods. {\color{black} The validity is rather determined 
by the importance and accuracy of the gas/wall treatment, 
as well as non-local effects in energy space.
In this respect a (properly defined) local Knudsen number
may be used. 
While a definition of the flow regime determined by the 
mass density gradient as proposed by Bird \cite{bird1994}
cannot always be applied (e.g., for incompressible flows), the gradient based definition 
with an appropriate choice of flow property (e.g., 
momentum), or a definition based on geometric 
considerations are reliable measures of the flow 
regime. Both definitions have been specified earlier. To illustrate this the local Knudsen number 
calculated from the momentum gradient of the DSMC 
results is plotted in figure \ref{fig:2d_knudsenNumber}. 
This gradient based Knudsen number profile clearly highlights 
the problematic spots, providing detailed spatially 
resolved information about the flow regime.
Additionally, as can be seen from the range of values, there is very good quantitative agreement between the geometric estimate and the gradient based calculation.
However, unlike geometric considerations, the gradient based definition can only be 
applied \textit{after} simulation results have been 
obtained, not \textit{before}. Consequently -- with a 
careful choice of the flow property $Q$ -- both 
definitions may be readily used to determine the 
flow regime. When a transitional or even rarefied gas 
flow is found}, the correct physics can only be 
captured by means of {\color{black} a kinetic model 
approach. In such cases continuum methods are 
vaguely} appropriate, only if the global flow behavior 
is of main interest, but not the detailed flow 
characteristics at small geometric features.

\section{Conclusions}\label{sec:conclusion}

In the present work the neutral gas flow under 
process conditions in the industrial scale PVD 
reactor {\color{black} CC800/9} Custom is analyzed. 
The validity of {\color{black} a continuum model} in 
the transition flow regime, characterized by a 
Knudsen number ranging between {\color{black} 
$0.05<\textit{Kn}<1.5$ (based on geometric considerations)}, is investigated. In 
particular, the analysis was accomplished using the 
{\color{black} CFD} simulation tool \textit{FLUENT}, 
as well as a modified version of the kinetic 
simulation tool \textit{dsmcFoam}. While the results 
from the continuum model remain approximately valid 
only up to a Knudsen number {\color{black} 
$\textit{Kn}<0.1$}, the kinetic model is able to 
predict the physically correct gas flow in {\color{black} all flow regimes, also} 
with larger Knudsen numbers of {\color{black} 
$\textit{Kn} > 0.1$}. The kinetic approach provides 
a more precise description especially at small 
geometric features. In addition, it can be easily 
extended to particular model requirements. With respect 
to the modeling of {\color{black} deposition processes (e.g., DC-MS or HPPMS)},
the sputtering process can be modeled, e.g., 
by providing an initial Thompson energy distribution 
paired with a cosine-law angular distribution {\color{black} 
for sputtered particles} \cite{kersch1994, serikov1996, 
malaurie1996, kadlec2007}. Moreover, for a deeper 
physical understanding, the kinetic model can be coupled 
to ``effective'' plasma models, or to a kinetic description 
of the ions, allowing for more complex and more complete 
energy exchange processes and chemical reactions. 
Consequently, it allows for a detailed analysis of the 
coating formation on the substrates in industrial, as 
well as research PVD processes. It can be stated that 
only kinetic simulations are able to appropriately merge 
the physics of the gas flow, the particle/wall 
interactions and the particle/particle interactions in 
sputtering processes. The development of a particle based 
kinetic simulation of all heavy particles {\color{black}-- 
neutrals and ions --} similar to the particle in cell 
(PIC) method \cite{birdsall1991, turner2013} will be the topic of a later study.

\section*{Acknowledgments}
The authors acknowledge the work of the 
\textit{OpenFOAM} community 
and in particular the authors of \textit{dsmcFoam}. 
This work is supported by the German Research Foundation 
(DFG) within the Collaborative Research Centre TRR 87 ``Pulsed
High Power Plasmas for the Synthesis of Nanostructured 
Functional Layers'' subproject C6.

\newpage

\listoffigures

\newpage

\clearpage

\clearpage

\clearpage

\clearpage

\clearpage

\clearpage

\newpage

\begin{thebibliography}{}

\bibitem{theiss2010}
S. Theiß, N. Bibinov, N. Bagcivan, M. Ewering, 
P. Awakowicz, K. Bobzin, 
J. Phys. D: Appl. Phys \textbf{43} (2010) 075205.
\bibitem{bobzin2008}
K. Bobzin, N. Bagcivan, P. Immich, S. Bolz, 
R. Cremer, T. Leyendecker, 
Thin Solid Films \textbf{517} (2008) 1251.
\bibitem{sanders2000}
{\color{black}D.M. Sanders, A. Anders, Surf. Coat. 
Tech. \textbf{133-134} (2000) 78.}
\bibitem{bird1994}
G.A. Bird, ``Molecular Gas Dynamics and the Direct 
Simulation of Gas 
Flows'', Oxford University Press, New York, USA, 1994.
\bibitem{bird1983}
G.A. Bird, Phys. Fluids \textbf{26} (1983) 3222.
\bibitem{boyd1995}
{\color{black}I.D. Boyd, G. Chen, G.V. Candler, Phys. 
Fluids \textbf{7} (1995) 210.}
\bibitem{anderson1995}
J.D. Anderson, ``Computational Fluid Dynamics: The Basics With 
Applications'', McGraw-Hill Science, New York, USA, 1995.
\bibitem{ferziger2002}
J.H. Ferziger, M. Peri\'c, ``Computational Methods for Fluid 
Dynamics'', 3rd edition, Springer, Berlin, Germany, 2002.
\bibitem{white2013}
C. White, M.K. Borg, T.J. Scanlon, J.M. Reese, Computers and 
Fluids \textbf{71} (2013) 261.
\bibitem{darbandi2011}
M. Darbandi, E. Roohi, Microfluid Nanofluid 
\textbf{10} (2011) 321.
\bibitem{bird1978}
G.A. Bird, Ann. Rev. Fluid Mech. \textbf{10} (1978) 11.
\bibitem{sazhin1997}
S.S. Sazhin, V.V. Serikov, Planet. Space Sci. \textbf{45} 
(1997) 361.
\bibitem{sun2004}
Q. Sun, I.D. Boyd, G.V. Chandler, J. Comp. Phys \textbf{194} 
(2004) 256.
\bibitem{zheng2006}
Y. Zheng, J.M. Reese, H. Struchtrup, J. Comp. Phys \textbf{218} 
(2006) 748.
\bibitem{wu2006}
J.-S. Wu, Y.-Y. Lian G. Cheng, R.P. Koomullil K.-C. Tseng, 
J. Comp. Phys. \textbf{219} (2006) 579.
\bibitem{kolobov2007}
V.I. Kolobov, R.R. Arslanbekov, V.V. Aristov, A.A. Frolova, 
S.A. Zabelok, J. Comp. Phys. \textbf{223} (2007) 589.
\bibitem{fluent2012}
ANSYS $\textregistered$ Academic Research, Release 14.0, 2012.
\bibitem{scanlon2010} 
T.J. Scanlon, E. Roohi, C. White, M. Darbandi, J.M. Reese, 
Computers and Fluids \textbf{39} (2010) 2078.
\bibitem{macpherson2009}
G.B. Macpherson, N. Nordin, H.G. Weller, Commun. Numer. Meth. 
Engng. \textbf{25} (2009) 263.
\bibitem{openfoam2012} 
OpenFOAM: the open source CFD Toolbox, User Guide 
version 2.1.1, 2012.
\bibitem{chapman1952}
{\color{black} S. Chapman, T.G. Cowling, ``The Mathematical 
Theory of Non-Uniform
Gases'', 2nd edition, Cambridge University Press, 
Cambridge, UK, 1952.}
\bibitem{thompson1968}
{\color{black} M.W. Thompson, A.E.R.E Harwell, 
Phil. Mag. \textbf{18} (1968) 377.}
\bibitem{sigmund1969}
{\color{black} P. Sigmund, Phys. Rev. \textbf{184} (1969) 383.}
\bibitem{mussenbrock2012} 
T. Mussenbrock, Contrib. Plasma Phys. \textbf{52} (2012) 571.
\bibitem{berg2005}
S. Berg, T. Nyberg, Thin Solid Films \textbf{476} (2005) 215.
\bibitem{berg1987}
S. Berg, H.-O. Blom, T. Larsson, C. Nender, J. Vac. 
Sci. Technol. A 
\textbf{5} (1987) 202.
\bibitem{depla2007}
{\color{black} D. Depla, S. Heirwegh, S. Mahieu and R. De Gryse, J. 
Phys D:
Appl. Phys. \textbf{40} (2007) 1957.}
\bibitem{kouznetsov1999}
V. Kouznetsov, K. Mac\'ak, J.M. Schneider, U. Helmersson, I. 
Pretrov, 
Surf. Coat. Tech. \textbf{122} (1999) 290.
\bibitem{lundin2009}
D. Lundin, N. Brenning, D. J\"adern\"as, P. Larsson, E. Wallin, 
M. Lattemann, M.A. Raadu, U. Helmersson, Plasma Sources Sci. 
Technol. \textbf{18} (2009) 045008.
\bibitem{gudmundsson2012}
J.T. Gudmundsson, N. Brenning, D. Lundin, U. Helmersson, J. Vac. 
Sci. Technol. A \textbf{30} (2012) 030801.
\bibitem{sarakinos2010}
{\color{black} K. Sarakinos, J. Alami, S. Konstantinidis, Surf.
Coat. Tech.
\textbf{204} (2010) 1661.}
\bibitem{kersch1994}
A. Kersch, W. Morokoff, Chr. Werner, J. Appl. Phys \textbf{75} 
(1994) 2278.
\bibitem{serikov1996}
V.V. Serikov, K. Nanbu, J. Vac. Sci. Technol. A \textbf{14} 
(1996) 3108.
\bibitem{malaurie1996}
A. Malaurie, A. Bessaudou, Thin Solid Films \textbf{286} (1996) 
305.
\bibitem{kadlec2007}
S. Kadlec, Plasma Process. Polym. \textbf{4} (2007) S419.
\bibitem{lundin2013}
{\color{black} D. Lundin, C. Vitelaru, L. de Poucques, N. Brenning, 
T. Minea, J. Phys. D: Appl. Phys. \textbf{46} (2013) 175201.}
\bibitem{mahieu2006}
{\color{black} S. Mahieu, G. Buyle, D. Depla, S. Heirwegh, P. 
Ghekiere, R. De Gryse, Nucl. Instr. and Meth. in Phys. Res. B. 
\textbf{243} (2006) 313.}
\bibitem{aeken2008}
{\color{black} K. Van Aeken, S. Mahieu, D. Depla, J. Phys D: Appl. 
Phys. \textbf{41} (2008) 205307.}
\bibitem{lugscheider2005}
{\color{black} E. Lugscheider, K. Bobzin, N. Papenfuß-Janzen, D.
Parkot, Surf. Coat. Tech. \textbf{200} (2005) 913.}
\bibitem{huo2012}
{\color{black} C. Huo, M.A. Raadu, D. Lundin, J.T. Gudmundsson, A. 
Anders, N. Brenning, Plasma Sources Sci. Technol. \textbf{21} 
(2012) 045004.}
\bibitem{yakhot1992}
V. Yakhot, S.A. Orszag, S. Thangam, T.B. Gatski, C.G. Speziale, 
Phys. Fluids A \textbf{4} (1992) 1510.
\bibitem{reif1987}
F. Reif, ``Fundamentals of Statistical and Thermal Physics'', 
McGraw-Hill, New York, USA, 1965.
\bibitem{birdsall1991}
C.K. Birdsall, IEEE Trans. Plasma Sci. \textbf{19} (1991) 65.
\bibitem{turner2013}
M.M. Turner, A. Derzsi, Z. Donk\'o, D. Eremin, S.J. Kelly, 
T. Lafleur, T. Mussenbrock, Phys. Plasmas \textbf{20} (2013) 
013507.

\end{thebibliography}
\end{document}